\documentclass[aps,prd,11pt,amsmath,amssymb,reprint,nofootinbib,showkeys,superscriptaddress]{revtex4-1} 

\usepackage{amssymb}
\usepackage{amsmath}
\usepackage{amsfonts}
\usepackage{graphicx}
\usepackage{color}
\usepackage{xspace}
\usepackage{ulem}
\usepackage{mathtools}
\usepackage{hhline}
\usepackage{tikz}

\usepackage{dcolumn}
\usepackage{bm}
\usepackage{hyperref}
\usepackage{comment}

\newcommand{\AddrSlovenia}{Jožef Stefan Institute, Jamova 39, 1000 Ljubljana,  Slovenia}
\newcommand{\AddrCoimbra}{Univ Coimbra, Faculdade de Ci\^encias e Tecnologia da Universidade de Coimbra and CFisUC, Rua Larga, 3004-516 Coimbra, Portugal}

\begin{document}

\title{Lukewarm inflation and the QCD axion}

\author{Paulo B. Ferraz}\email{paulo.ferraz@student.uc.pt}\affiliation{\AddrCoimbra}
\author{Ant\'onio Torres Manso}\email{antonio.torres.manso@ijs.si}\affiliation{\AddrCoimbra}\affiliation{\AddrSlovenia}
\author{Jo\~{a}o G.~Rosa} \email{jgrosa@uc.pt}\affiliation{\AddrCoimbra}

\date{\today}

\begin{abstract}

We show that a late period of warm inflation, at temperatures just above the QCD scale, can dilute the abundance of the QCD axion produced by the misalignment mechanism, thus avoiding the need to fine-tune the initial misalignment angle in scenarios with an axion decay constant close to the GUT scale. We develop a concrete realization of this lukewarm inflation stage involving right-handed neutrinos in $0.1-1$ GeV mass range and an associated scalar sector, showing that coherent axion oscillations are damped during this period. In this scenario, a previously generated large baryon asymmetry may also be diluted to yield the observed value, and lukewarm inflation is followed by an early-matter era dominated by the lightest right-handed neutrino, before it decays into Standard Model states.
This model can be tested via changes to the number of relativistic species and the induced small-scale enhancements of the primordial curvature power spectrum, leading to primordial black holes and scalar-induced gravitational waves.
\end{abstract}


\maketitle

\section{Introduction}

In the standard cosmological paradigm, an early period of accelerated expansion, known as inflation, is the most accepted explanation for the homogeneity and flatness of the observable universe and for the almost scale invariance of primordial curvature perturbations inferred from CMB and Large-Scale Structure observations \cite{infGuth,inflinde,infalbrecht}. This period of inflation, which precedes the hot Big Bang evolution, is typically driven by a scalar field, known as the inflaton, that undergoes a period of ``slow-roll'' evolution. Fluctuations in this quantum scalar field then naturally lead to a nearly scale-invariant spectrum of scalar curvature perturbations on super-horizon scales.

Nevertheless, the thermal history of the Universe prior to Big Bang Nucleosynthesis (BBN) remains largely unconstrained, and several open problems persist at the interface of inflationary cosmology and particle physics. These include the origin of the dark matter abundance \cite{Cirelli:2024ssz}, the smallness of the baryon asymmetry \cite{Cline:2006ts}, and the origin of the neutrino masses \cite{Hernandez:2025spl}. 

A popular dark matter candidate is the QCD axion, first introduced to solve the strong CP-problem \cite{Peccei:1977hh,Preskill:1982cy,Preskill:1982cy}. In the standard misalignment scenario, which assumes that the Peccei-Quinn symmetry is spontaneously broken before the last 60 e-folds of inflation, the axion field is initially displaced from the minimum of its QCD-induced potential and undergoes coherent oscillations about it. However, the QCD axion may account for a fraction of the dark matter in the universe only for decay constants $f_a \lesssim 10^{12}$ GeV if the initial misalignment angle $\theta_i \lesssim \mathcal{O}(1)$ \cite{Marsh:2015xka, OHare:2024nmr}. Larger values of $f_a$, including the theoretically appealing GUT-scale regime $f_a \sim 10^{15\text{--}16}$ GeV, generically lead to axion overproduction and require either finely tuned initial conditions or modifications of the cosmic thermal history. Light relics such as the QCD axion are particularly sensitive to any departure from the standard post-inflationary thermal history, since their abundance depends on both the expansion rate and late-time entropy production \cite{Steinhardt:1983ia,Lazarides:1990xp}.

 A broad class of alternatives has therefore been explored in the literature to dilute the axion relic density through late-time entropy production or non-standard cosmological epochs. Early examples include entropy release from heavy particle decays \cite{Kawasaki:2011ym} or moduli domination \cite{Kawasaki:1995vt}, thermal inflation \cite{Hattori:2015rga}, low scale inflation \cite{Graham:2018jyp,Takahashi:2018tdu}, and additional late periods of accelerated expansion \cite{Hoof:2017ibo}, which can reduce the axion abundance by increasing the comoving entropy density after the onset of axion oscillations. 
 
While these mechanisms are phenomenologically viable, they typically face at least one of the following challenges:
(i) they require introducing additional ad hoc scalar fields or moduli with finely arranged decay rates;
(ii) they rely on inflation followed by extremely low reheating temperatures, which are difficult to achieve without spoiling reheating efficiency or BBN consistency;
(iii) they dilute all relics uniformly, often erasing any pre-existing baryon asymmetry unless baryogenesis is re-engineered at very low scales;
(iv) they require tuning the inflaton and axion masses such that $m_\phi \sim H \sim m_a$ to affect the onset of axion oscillations.

In this work, we propose a novel mechanism for axion dilution based on a secondary, late stage warm inflation period, which we denote as {lukewarm inflation}. Warm inflation generically  features continuous energy transfer from the inflaton to a thermal bath during inflation, leading to sustained radiation and entropy production throughout the inflationary epoch \cite{Berera:1995ie, Berera:2008ar,Bartrum:2013fia}. As we show, this framework can naturally accommodate low-scale inflation without the need for a separate reheating phase and naturally provides a graceful exit into radiation domination.

There is no fundamental reason to consider a single-stage inflation scenario, with a single scalar field driving the required $50-60$ e-folds of accelerated expansion. In fact, most single field scenarios require large inflaton energy densities to consistently explain the observed amplitude of primordial curvature perturbations, which inevitably leads to large temperatures in the post-inflationary universe. This could potentially result in an overproduction of thermal relics in addition to the QCD axion, such as e.g.~gravitinos in supersymmetric models or monopoles in the case of high-temperature phase transitions that could overclose the universe. One or more stages of secondary inflation periods could significantly dilute the abundance of these relics down to presently unobservable values. Theoretically, one may also argue that most extensions of the Standard Model (like supersymmetry, extra-dimensions or ultimately superstring/M-theory) include a plethora of scalar fields that, through different dynamical realizations, could temporarily dominate the universe's energy balance as approximate cosmological constants at different epochs (see e.g.\cite{Ferraz:2024bvd}).

The lukewarm inflationary phase that we propose in this work is characterized by strong dissipative effects for less than 10 e-folds, a Hubble scale comparable to the axion mass and a slowly decreasing temperature $T \lesssim1$ GeV. 
This scenario has several appealing features compared to other dilution mechanisms:
(i) the QCD axion starts oscillating as cold dark matter during lukewarm inflation,  a period in which entropy is continuously produced, leading to an efficient dilution of its abundance;
(ii) any pre-existing baryon asymmetry and other thermal relics are more diluted than the QCD axion;
(iii) the Universe exits lukewarm inflation directly into a low-temperature radiation bath;
(iv) lukewarm inflation with $H \sim m_a$ can be realized without finely tuning the inflaton and axion masses;
(v) the dissipative dynamics can be connected directly to the Standard Model and the neutrino sector in particular, providing a direct connection between axion cosmology, lukewarm inflation dynamics and neutrino mass generation \cite{Levy:2020zfo}.

In this paper we show that a QCD axion with ${f_a \simeq 10^{15\text{--}16}\,\mathrm{GeV}}$ and a typical initial misalignment angle $\theta_i \sim \mathcal{O}(1)$ can account for the entirety of the observed dark matter abundance, considering a short lukewarm inflationary epoch in the strong dissipative regime. The same dynamics also dilutes an initially $\mathcal{O}(1)$ baryon asymmetry to its observed value today while remaining consistent with BBN and current bounds on $\Delta N_{\rm eff}$. We identify the viable parameter space, present the full cosmological evolution of the axion–inflaton–radiation system, and discuss the potential observational prospects of the model.

This work is organized as follows. In the next section, we revise the basic characteristics of the QCD axion and its evolution throughout the cosmic history. In Section \ref{sec:LukewarmI}, we describe the main features of the  lukewarm inflation model, detailing the particle content and relevant interactions. In Section \ref{sec:LukewarmII} we study the dynamics of the model, following the evolution of the different components. We determine in particular the parametric regimes with a viable QCD axion abundance and baryon asymmetry, as well as a successful transition into the standard radiation era before BBN. In Section \ref{sec:Observ} we discuss some of the observational consequences of lukewarm inflation, namely the prediction of dark radiation and enhancement of small-scale curvature perturbations. We conclude in Section \ref{sec:concl} by summarizing the main features of our model and discussing prospects for future work.

\section{Axion Cosmology}


The QCD axion mass can be written in terms of its decay constant $f_a$, from chiral perturbation theory \cite{GrillidiCortona:2015jxo} and numerical lattice simulations \cite{Borsanyi:2016ksw}, as:
\begin{align}
    m_a = 5.7\left(\frac{10^{12}\text{ GeV}}{f_a}\right)\ \mu\text{eV}.
\end{align}
When taking thermal effects into account the axion mass becomes temperature-dependent,
\begin{align}
    m_a^2(T) =
    \begin{cases}
        m_a^2\left(\frac{T}{T_{QCD}}\right)^{-n},\quad &T>T_{QCD}\\
        m_a^2\quad,\quad  &T<T_{QCD},\label{Eq:Axionmass}
    \end{cases}
\end{align}
where $T_{QCD}$ defines the temperature at which non-perturbative QCD effects become significant and $n\simeq 8$ in the dilute instanton gas approximation \cite{Gross:1980br,Wantz:2009it,Borsanyi:2016ksw}.

In the standard misalignment mechanism, the axion field is set at an initial value ${a_i=f_a\theta_i}$ defined by an arbitrary initial angle $\theta_i$ that generically is different from the minima of the axion potential. The value of the initial angle depends on the dynamics of Peccei-Quinn symmetry breaking, i.e.~whether it is broken before or after inflation. In the pre-inflation scenario, the misalignment angle $\theta_i$ takes a random value in the Hubble-size patch that inflates to eventually become the presently observable universe. Therefore, we expect that statistically
$|\theta_i| \sim 1$. If, however, symmetry breaking occurs after inflation, several disconnected patches with different values of $\theta_i$ become causally connected after inflation. Then, a natural value for the initial angle is the average over all of its possible values $\langle \theta_i\rangle\simeq \sqrt{\langle\theta_i^2\rangle}\simeq \pi/\sqrt{3}$. In both cases, a value of the initial angle $\theta_i\sim1$ seems to be natural, as we will consider for the remainder of this work.

From the initial conditions for the axion, $\theta_i \sim 1$ and $\dot{\theta}_i \simeq 0$, and in the regime where $T\gg \Lambda_{\text{QCD}}$ and $H\ll m_a(T)$, the amplitude of the axion field is practically frozen at $\theta\simeq \theta_i$. Once the temperature goes below the QCD scale and the Hubble parameter becomes comparable to the axion mass, the axion starts oscillating and behaves as pressureless matter:
\begin{align}
    \rho_a(t) \simeq\rho_a(t_{osc})\left(\frac{a(t_{osc})}{a(t)}\right)^3,\quad t>t_{osc},
\end{align}
where $t_{osc}$ is the time at which the axion field starts to oscillate coherently. Moreover, the axion energy density when it starts oscillating can be approximated by
\begin{align}
    \rho_a(t_{osc})\simeq \frac{m_a^2}{2}f_a^2\langle\theta_i^2\rangle~,
\end{align}
up to sub-leading anharmonic effects.
Assuming the standard cosmological evolution, the present QCD axion abundance is then given by \cite{Marsh:2015xka,OHare:2024nmr}
\begin{align}
    \Omega_ah^2 \equiv \frac{\rho_{a,0}}{\rho_{c,0}} \simeq 0.12\left(\frac{\theta_i}{0.86}\right)^2\left(\frac{f_a}{10^{12}\text{ GeV}}\right)^{1.16},
\end{align}
with $\rho_{c,0}=3H_0^2M_p^2=8.07  \times 10^{-11} h^2 \mathrm{eV}^4$.
Thus, for typical values of the initial misalignment angle, ${f_a\lesssim 10^{12}\text{ GeV}}$ or otherwise the QCD axion would overclose the universe. One way to allow for larger values of $f_a$ is to introduce a new source of entropy production such that the ratio $n_a/s$ decreases. Earlier works along these lines are able to accommodate values up to $f_a\sim 10^{15}$ GeV \cite{Steinhardt:1983ia,Lazarides:1990xp,Kawasaki:1995vt,Kawasaki:2011ym,Hattori:2015rga,Hoof:2017ibo,Takahashi:2018tdu,Graham:2018jyp}. 

In our model, there are, in fact, two distinct entropy sources of entropy, firstly the dissipative entropy production during the lukewarm inflation period itself and, secondly, entropy resulting from the post-lukewarm inflation decay of the right-handed neutrinos produced by the dissipative effects. As we discuss below, these right-handed neutrinos may drive an early matter-dominated epoch just after the lukewarm inflation stage, before decaying and ``reheating'' the Standard Model degrees of freedom.

We may therefore write the final axion abundance as:
\begin{equation} \label{axion_abundance_final}
\Omega_{a,0} = \Delta_s \left({a_{osc}\over a_E}\right)^3{s_{osc}\over s_E}\Omega_{a,osc}~,
\end{equation}
where the subscripts ``osc'' and ``E'' denote the time at which axion oscillations begin and lukewarm inflation ends, respectively, and $\Delta_s$ is the dilution factor due to right-handed neutrino decays. Thus, we see that if the QCD axion starts to oscillate during the lukewarm inflation period, such that $s_{osc}\simeq s_E$, its abundance can be efficiently diluted. 
In the next sections we describe the lukewarm inflation dynamics and the underlying particle physics model, and perform a detailed study of the evolution of the QCD axion abundance during this period.

\section{Lukewarm inflation model}\label{sec:LukewarmI}

Warm inflation is, in general, characterized by thermal effects that result from interactions between the inflaton field and the particles in a sub-dominant nearly-thermal bath \cite{Berera:1995ie,Berera:1995wh,Berera:2008ar,Bartrum:2013fia}. Alongside damping its motion, these interactions dissipate the inflaton's energy into the thermal bath, preventing the latter's exponentially fast dilution from accelerated expansion. Dissipative effects are encapsulated in the thermal dissipation coefficient $\Upsilon$, which leads to a local friction term $\Upsilon\dot\phi$ in the inflaton's equation of motion, at least in the adiabatic regime where the inflaton evolves slowly compared to the response of the thermal bath, and to a corresponding radiation source term. 
The form of the dissipation coefficient $\Upsilon$ depends on the microphysics and may lead to different realizations of warm inflation \cite{Bastero-Gil:2009sdq,Bastero-Gil:2010dgy,Bastero-Gil:2012akf}. 

In this work, we consider the Warm Litle Inflaton (WLI) model \cite{Bastero-Gil:2016qru}, where the inflaton is a singlet scalar field resulting from the collective spontaneous breaking of a $U(1)$ gauge symmetry. Specifically, the scenario considers two complex scalar fields $\Phi_1$ and $\Phi_2$ with the same $U(1)$ charge $q$ and vacuum expectation value $\langle |\Phi_1|\rangle = \langle |\Phi_2|\rangle = M/\sqrt{2}$. This is ensured by an underlying discrete interchange symmetry, as we describe below. The vacuum manifold may thus be parametrized as
\begin{align}
    \Phi_1 = \frac{M}{\sqrt{2}}e^{i(\alpha+\phi)/M},\qquad \Phi_2 = \frac{M}{\sqrt{2}}e^{i(\alpha-\phi)/M}.
\end{align}
The global phase $\alpha$ corresponds to the Nambu-Goldstone (NG) boson of the spontaneously broken symmetry, which is incorporated as the longitudinal component of the massive gauge boson in the unitary gauge. The relative phase $\phi$ is the remaining physical scalar degree of freedom after symmetry breaking and, since it is gauge invariant, we may include arbitrary contributions to the scalar potential $V(\phi)$ in the Lagrangian.

The two complex scalar fields are coupled, via Yukawa terms, to two additional fermions $\psi_1$ and $\psi_2$ that, alongside their decay products, make up the thermal bath during warm inflation. While the left-handed components of each fermion are taken to have the same charge as the complex scalar fields, their right-handed counterparts are $U(1)$-singlets. Furthermore, as mentioned above, a discrete interchange symmetry $\Phi_1 \leftrightarrow i\Phi_2$ and $\psi_1 \leftrightarrow \psi_2$ is imposed, yielding the Lagrangian interaction\footnote{Note that this is not the most general form of the interaction Lagrangian \cite{Levy:2020zfo, Gorgulho:2025wxz}, although we consider it for simplicity and without loss of generality.}:
\begin{align}
    -\mathcal{L}_{\phi\psi} &= \frac{g}{\sqrt{2}}(\Phi_1+\Phi_2)\overline{\psi}_{1L}\psi_{1R}-i\frac{g}{\sqrt{2}}(\Phi_1-\Phi_2)\overline{\psi}_{2L}\psi_{2R}\\
    &=gM\cos\left(\frac{\phi}{M}\right)\overline{\psi}_{1L}\psi_{1R}+gM\sin\left(\frac{\phi}{M}\right)\overline{\psi}_{2L}\psi_{2R}.
\end{align}
The fermion masses are therefore oscillatory functions of the inflaton field and, hence, bounded $|m_{1,2}| \le gM$. If the temperature during inflation takes the values $gM \lesssim T\lesssim M$, the fermions are thus light degrees of freedom and thus contribute significantly to the entropy. These interactions introduce thermal corrections to the scalar potential, but since the leading contribution $\Delta V_T \propto (m_1^2+m_2^2)T^2$ (up to numerical factors) is independent of the inflaton field, it does not contribute to the eta slow-roll  parameter, see equation \eqref{eq:slowroll}. In other words, this mechanism ensures the cancellation of the leading thermal corrections to the inflaton mass that could potentially prevent a slow-roll evolution. We note, however, that warm inflation can be realized in this setup even with only a single fermion species, given the oscillatory nature of the induced thermal corrections to the scalar potential, as shown in \cite{Ferraz:2023qia}.

In addition, the WLI model includes Yukawa interactions that allow the fermions $\psi_{1,2}$ to decay into another chiral fermion $\psi_\sigma$ and a scalar $\sigma$, with appropriately chosen $U(1)$ charges:
\begin{align}
    -\mathcal{L}_{\psi\sigma} = h\sigma\sum_{i=1,2}\left(\overline{\psi}_{iL}\psi_{\sigma R}+\overline{\psi}_{\sigma L}\psi_{i R}\right),
\end{align}

The dissipation coefficient $\Upsilon$ may thus be computed using standard non-equilibrium field theory techniques \cite{Kapusta:2006pm}, and in the high-temperature regime $T\gtrsim m_{1,2}$ takes the form:
\begin{align}
    \Upsilon \simeq C_T T,\quad C_T=\frac{g^2}{h^2}\frac{3}{1-0.34\log h}.\label{eq: dissipation coefficient}
\end{align}

For the lukewarm inflation period, we will consider a specific realization of this generic WLI setup, where similarly to the scenario proposed in \cite{Levy:2020zfo}, the fermions $\psi_{1,2}$ are identified with two of the right-handed neutrinos $N_{1,2}$. The additional fermion $\psi_\sigma$ (corresponding to a lepton in \cite{Levy:2020zfo}) will here be identified with the third right-handed neutrino $N_3$ in a minimal extension of the Standard Model accounting for small neutrino masses through the type I seesaw mechanism. This means that $N_1$ and $N_2$ are degenerate after lukewarm inflation, when the inflaton settles at the minimum of its potential. We assume that $N_3$ is the lightest right-handed neutrino, although its mass is comparable to the other two. 

Let us note that this identification requires replacing the Dirac-type Yukawa terms described above by analogous Majorana-terms, as described in \cite{Levy:2020zfo}. However, the resulting dissipation coefficient only differs from Eq.~(\ref{eq: dissipation coefficient}) by numerical factors that can be absorbed into effective Yukawa couplings $g$ and $h$, so that we will not include them here explicitly. 

We assume, for simplicity, that the additional scalar $\sigma$ has a zero vacuum expectation value, noting that $\langle \sigma\rangle \neq 0$ would lead to Majorana mass mixing in the right-handed neutrino sector. Although its exact mass value does not affect the lukewarm inflation dynamics as long as it is below the temperature of the radiation bath, we will consider a scenario where it is light enough to behave as dark radiation at late times.

The three right-handed neutrinos generate the observed light neutrino masses through the usual Yukawa Higgs interactions, which therefore link the lukewarm inflationary dynamics to the Standard Model:
\begin{equation}
	\mathcal{L}_{\text {Yuk }}=\sum_{i, j}\left(-Y_{i j} \bar{N}_i H^{\dagger} L_j+\text { h.c. }\right),
\end{equation}
where $H$ field is the Higgs, $L_i$ are the lepton doublets and $Y$ is the Yukawa couplings matrix.

As we discuss below, a lukewarm inflation period with a Hubble parameter $H\sim m_a$ and $f_a\sim 10^{15}-10^{16}$ GeV requires right-handed neutrinos with masses  $M_N\sim1$ GeV, for which the partial decay width into Standard Model particles is given by
\begin{equation}
	\Gamma_{N_i} \simeq \frac{G_F^2 M_i^5}{96 \pi^3} \sum_\alpha\left|U_{\alpha i}\right|^2,
\end{equation}
where the mixing matrix is defined as 
\begin{equation}
	\left|U_{\alpha i}\right|^2=\frac{\left|m_{D\ \alpha i}\right|^2}{M_i^2}=\left\langle v\right\rangle^2\frac{\left|Y_{\alpha i}\right|^2}{M_i^2},
\end{equation}
and $\langle v\rangle$ denotes the Higgs vacuum expectation value.
The decay rate may thus be written as 
\begin{equation}
	\Gamma_{N_i} \simeq \frac{G_F^2 \left\langle v\right\rangle^2 M_i^3}{96 \pi^3} \sum_\alpha\left|Y_{\alpha i}\right|^2.
\end{equation}

We may relate this to the light neutrino masses generated via the type I seesaw mechanism. As an estimate, let us consider the case of a single right-handed neutrino species, for which:
\begin{equation}
	m_\nu \simeq-m_D M^{-1} m_D^T\simeq  Y_{eff}^2\frac{v^2}{M_N}.
\end{equation}
For a light neutrino mass of $m_\nu\sim 0.05$ eV, the effective Yukawa coupling, $Y_{eff}^2=\sum_\alpha Y_{\alpha i}^2$, 
  takes the value:  
\begin{equation}
	Y_{eff}^2\simeq 1.65 \times 10^{-15} \frac{M_i}{\mathrm{GeV}}.\label{Eq:Seesaw}
\end{equation}

The smallness of this effective coupling implies that the right-handed neutrinos cannot efficiently decay into Standard Model particles during the lukewarm inflation period. This means that, during this secondary inflationary stage, only the three right-handed neutrinos and the additional scalar $\sigma$ are present in the thermal bath, and that all known particles are diluted away by the inflationary expansion (which as we describe below also has an impact on the dynamics of the QCD axion). 

We must therefore ensure that the right-handed neutrinos decay into Standard Model particles after lukewarm inflation and before BBN. The heaviest states $N_1$ and $N_2$ decay mainly into $N_3$ and $\sigma$ after lukewarm inflation ends and the temperature falls below their mass. Hence, we need to ensure that the lightest right-handed neutrino $N_3$ decays efficiently into leptons, which in turn thermally produce the remaining light degrees of freedom, before standard neutrino decoupling at around $T_{BBN}\sim 5$ MeV. The condition $\Gamma_{N_3}> H(T_{BBN})$ thus results in a lower bound on the right-handed neutrino masses at around a few hundred MeV, potentially as low as 100 MeV depending on the details of the neutrino mass generation mechanism \cite{Hernandez:2014fha,Chrzaszcz:2019inj, Domcke:2020ety,Abdullahi:2022jlv,Hernandez:2025spl}. 

In this right-handed neutrino mass range, it is easy to check that Standard Model degrees of freedom are not excited at the temperatures $T\sim 1$ GeV typical of the lukewarm inflation scenario that we are interested in. They are, however, thermally produced at temperatures above the electroweak scale, for which the right-handed neutrinos may directly decay into Higgs-lepton pairs. Hence, in the next section where we discuss the dynamics of lukewarm inflation, including the QCD axion, we will consider that before this secondary inflation stage the Standard Model degrees of freedom are in equilibrium with the right-handed neutrino sector, decoupling during lukewarm inflation and being reheated at late times, just before BBN.


\section{Lukewarm inflation dynamics}\label{sec:LukewarmII}

In this section, we consider the dynamics of a secondary lukewarm inflation scenario with a Hubble constant close to the mass of the QCD axion, such that the latter starts oscillating coherently about the minimum of its potential during this period.

As discussed above, the evolution of the scalar inflaton field driving this period is damped by the interactions with a thermal bath, including the three right-handed neutrinos and the additional light scalar $\sigma$, following an effective Langevin equation of the form \cite{Berera:1995wh,Berera:1999ws,Berera:2008ar,Bastero-Gil:2009sdq}:
\begin{equation} \label{langevin}
\ddot{\phi}+(3H+\Upsilon)\dot{\phi}+V_\phi=\xi,
\end{equation}
where $\xi$ is the thermal noise induced by dissipative friction, which close to thermal equilibrium is approximately a Gaussian white-noise term. Given that for now we are mainly interested in the evolution of the background inflaton field, we will drop this term since $\langle \xi\rangle =0$. For illustrative purposes we will consider a quadratic inflaton potential $V(\phi) = m_\phi^2\phi^2/2$, although other forms are possible given that the inflaton is invariant under gauge transformations\footnote{The discrete interchange symmetry is effectively a $\mathbb{Z}_2$ symmetry acting on the inflaton field up to a field redefinition, so imposing this symmetry implies that the potential must be an even function of the field.}.

For temperatures above the right-handed neutrino masses the thermal friction sources a nearly-thermal radiation bath with energy density $\rho_R={\pi^2\over 30}g_* T^4$ and entropy density $s=4\rho_R/3T$, such that:
\begin{equation}
\dot{\rho}_R+4H\rho_R=\Upsilon\dot\phi^2~.\\
\end{equation}

The dynamics of the QCD axion $a=\theta f_a$ is then determined by:
\begin{equation}
\ddot{\theta}+3H\dot{\theta}+m_a^2(T)\sin\theta=0~,    
\end{equation}
which, along with the Friedmann equation:
\begin{equation}
  H^2 = {\rho_\phi+\rho_R+\rho_a\over 3 M_P^2}~,
\end{equation}
completes the full set of differential equations that determine the dynamics of the relevant fields during the secondary lukewarm inflation period.


For $H\sim m_a$, we have $T/H\gtrsim \Lambda_{QCD}/m_a\sim f_a/\Lambda_{QCD}\gg1$, such that $\Upsilon \gg 3H$ and lukewarm inflation occurs in a (very) strong dissipation regime.

In the slow-roll regime, the inflaton and radiation equations can be written in the form:
\begin{align}\label{slow-roll}
&\frac{\phi'}{M_p}= -\frac{\sqrt{2\epsilon_\phi}}{1+Q},\\
&\frac{Q'}{Q} =\frac{6\epsilon_\phi-2\eta_\phi}{3+5Q},
\end{align}
where $Q={\Upsilon\over 3H}\gg 1$ and primes denote  derivatives with respect to the number of e-folds $N_e$ of lukewarm inflation and the slow-roll parameters satisfy the conditions:
\begin{align}
    \epsilon_\phi = \frac{M_P^2}{2}\left(\frac{V_\phi}{V}\right)^2\ll1+Q,\quad |\eta_\phi| = M_P^2\left|\frac{V_{\phi\phi}}{V}\right|\ll 1+Q~.\label{eq:slowroll}
\end{align}
For a quadratic scalar potential and $Q\gg1$, the slow-roll equations can be integrated to give:
\begin{align}
    &\phi = \phi_{I}\left(1-\alpha N_e\right)^{5/6},\\
    &T =T_{I}\left(1-\alpha N_e\right)^{1/6},\label{eq: slowrollsolT}\\
    &H = H_{I}(1-\alpha N_e)^{5/6},\label{eq: slowrollsolH}
\end{align}
where the subscript `I' labels the corresponding quantity at the start of the lukewarm inflation stage, i.e.~when the slow-roll conditions are first satisfied, and:
\begin{align}
    \alpha = \frac{12}{5}\left(\frac{M_P}{\phi_I}\right)^2Q_{I}^{-1} =\left({N_{LW}+{5\over 6}}\right)^{-1} ~.\label{eq: alphaSR}
\end{align}
The second equality in equation \eqref{eq: alphaSR} follows from the condition $\epsilon_H = -H'/H = 1$ at the end of lukewarm inflation and $N_{LW}$ denotes the total number of e-folds in this period. 

The dynamics of lukewarm inflation is, in this case (and analogously for other monomial potentials) completely determined by only 3 parameters, which we may choose as e.g.~the scale of lukewarm inflation $H_{I}$, its duration $N_{LW}$ and the dissipation constant $C_T$, which we recall corresponds essentially to the squared ratio of the Yukawa couplings defining the particle physics model. All other quantities may be expressed in terms of these using the slow-roll equations for the field and radiation energy density, which yield, in particular, for the initial radiation temperature:
\begin{equation}
T_{I}^4\simeq {75\over 2\pi^2g_*}\left(N_{LW}+{5\over 6}\right)^{-1}H_{I}^2 M_P^2~, 
\end{equation}
which we may rewrite as:
%

\begin{align}
T_{I}\simeq &~{1.6\over g_*^{1\over4}}\!\left({H_{I}\over m_a}\right)^{1\over2}\!
\left({10^{16}\ \mathrm{GeV}\over f_a}\right)^{1\over2}
\!\left(N_{LW}+{5\over 6}\right)^{-{1\over4}}\ \mathrm{GeV}.
\end{align}

This shows that a lukewarm inflation period with ${H_I\sim m_a}$ and an axion decay constant close to the GUT scale must occur at temperatures below the GeV scale, thus also setting an upper bound on the mass of the right-handed neutrinos. Given the lower bound discussed earlier, from requiring that the lightest right-handed neutrino decays into Standard Model particles before BBN, this sets the required right-handed neutrino mass scale around a few hundred MeV.

The inflaton mass is also determined by the slow-roll equations, yielding:
%
%

\begin{align}
m_\phi\simeq &~0.8{C_T^{1\over2}\over g_*^{1\over8}} \left( \frac{H_{I}}{ m_a}\right)^{3\over4}\!\left(\frac{10^{16} \mathrm{GeV} }{f_a}\right)^{3\over4}\!\left(N_{LW}+{5\over 6}\right)^{-{5\over8}}\;\mathrm{eV}.
\end{align}

This means that the scalar field driving the lukewarm inflation period must have a sub-eV mass, although $m_\phi \gg H_I \sim m_a$. Note that this is only possible as slow-roll is sustained by strong thermal friction, with Hubble friction playing a minor role in this regime. The corresponding field values are also well below the Planck scale, since from the Friedmann equation $\phi\sim (H/m_\phi) M_P\ll M_P$.

In the above discussion, $g_* = 7/8\times2\times3+1 =6.25$ refers to the number of light degrees of freedom in the thermal bath during lukewarm inflation, including the three right-handed neutrinos and the $\sigma$ scalar. Before the onset of lukewarm inflation, all Standard Model degrees of freedom are present in the radiation bath, but since neither dissipation nor other thermal processes involving the right-handed neutrinos can source them efficiently during the lukewarm inflation, they are simply diluted away by accelerated expansion.

\begin{figure}
    \centering
    \includegraphics[width=1\linewidth]{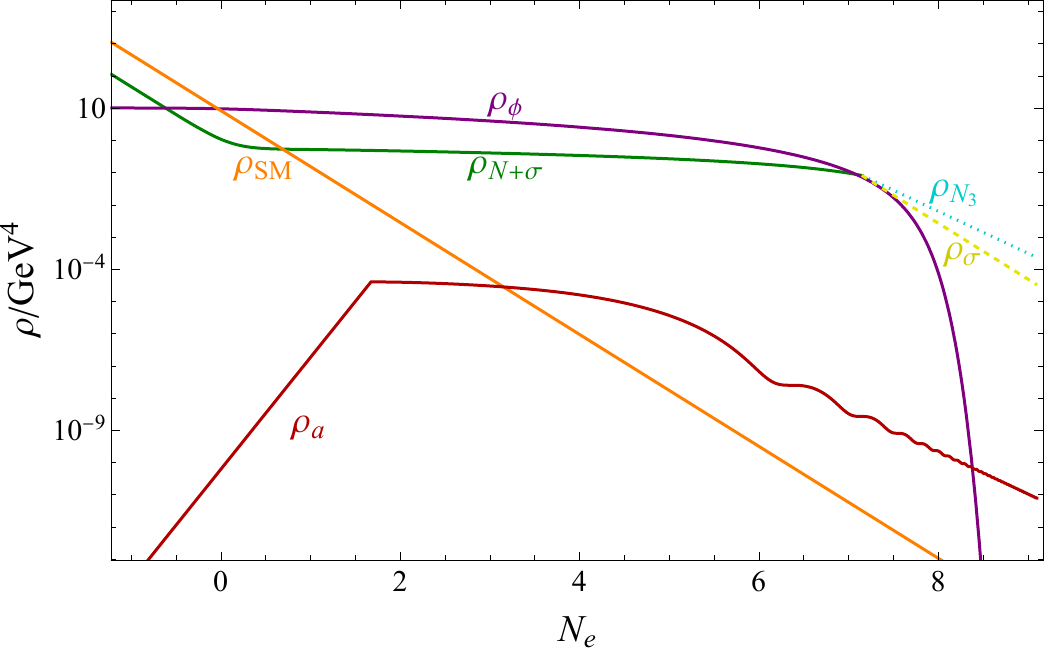}
    \caption{Evolution of the different components of the energy density during lukewarm inflation, for $m_a/H_I \simeq 0.55$, ${C_T\simeq 10^{-3}}$, $N_{LW} \simeq 7$ e-folds and $m_\phi \simeq 9$ meV. The axion decay constant is $f_a=10^{16}$ GeV, corresponding to $m_a\simeq 0.6$ neV, with an initial misalignment angle $\theta_i=\pi/\sqrt{3}$.}
    \label{fig: totaldyn1}
\end{figure}

In Figure \ref{fig: totaldyn1} we show the evolution of the energy densities for the different components of our model for an illustrative choice of parameters. As one can see in this figure, the system starts in a radiation dominated era with the Standard Model $\rho_{SM}$ and the right-handed neutrino sector (including $\sigma$) at the same temperature. We assumed that these components are the product of a reheating mechanism after the primary inflationary stage. Notice that the lukewarm inflaton is already following a slow-roll trajectory, which was made easier due to strong dissipation and it being a subdominant component (see e.g. \cite{Ferraz:2024bvd}). As the inflaton field becomes dominant and lukewarm inflation begins, the Standard Model light degrees of freedom simply redshift away as usual radiation, while the right-handed neutrino sector's energy density stabilizes at a sub-dominant value as a result of the thermal friction source. Note that the slow-roll equations imply $\rho_R/\rho_\phi \simeq \epsilon_H/2\simeq \epsilon_\phi /2Q$ for $Q\gg 1$, so that radiation is always sub-dominant in the slow-roll regime. Once the slow-roll conditions are violated, radiation smoothly takes over. In the present setup, as expected, the temperature falls below the mass of the lightest right-handed neutrino roughly at the end of the slow-roll period, so that it is diluted as cold matter, as opposed to the light $\sigma$-scalar, which behaves as dark radiation after the lukewarm inflation stage. This shows that in this setup lukewarm inflation is naturally followed by an early matter-dominated stage, starting at temperatures around a few hundred MeV. This short period ends with the decay of $N_3$ into Standard Model particles before BBN, although this is not shown in Figure \ref{fig: totaldyn1}.

\begin{figure}[!h]
    \centering
    \includegraphics[width=1\linewidth]{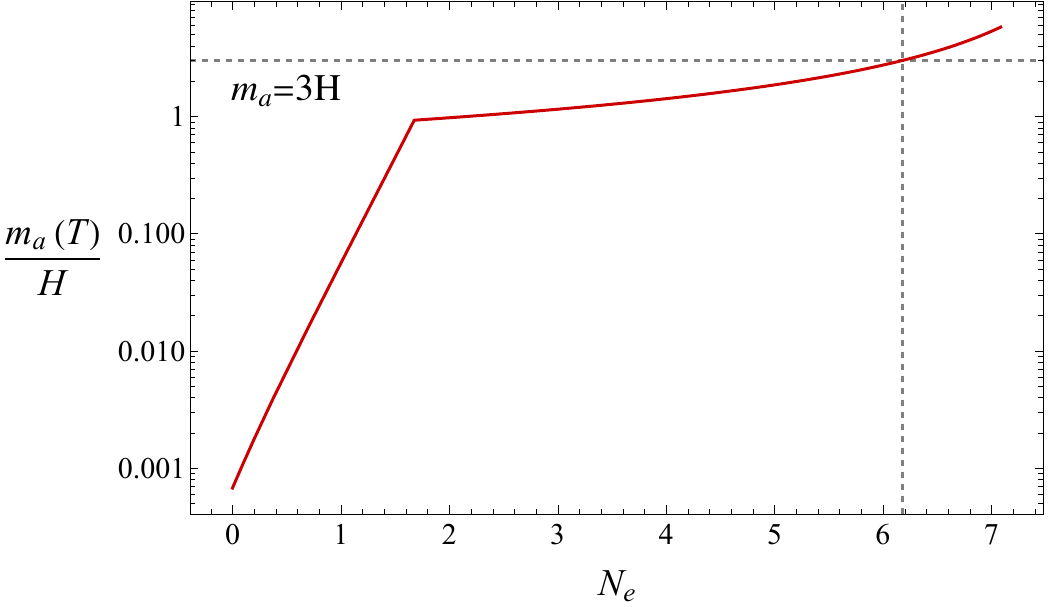}
    \caption{Evolution of the ratio between the QCD axion mass and the Hubble parameter, for the example shown in Figure \ref{fig: totaldyn1}.}
    \label{fig: axionmvsH}
\end{figure}

This figure also shows the evolution of the QCD axion's energy density during approximately 7 e-folds of lukewarm inflation. Let us first note that the axion mass depends on the temperature of the QCD degrees of freedom, which simply redshifts as $a^{-1}$, so that the axion mass reaches its zero-temperature value after only a couple of e-folds, as illustrated in \ref{fig: axionmvsH}. In this example, $m_a\simeq H_I$, and so the axion field is initially critically damped. This slow evolution damps the misalignment angle with which it begins to oscillate, even though this occurs only close to the end of the lukewarm inflation stage. 

To determine the final abundance of the QCD axion, we must determine the entropy produced in the decay of $N_3$ into Standard Model particles. Firstly, note that at the end of lukewarm inflation the temperature first falls below the mass of $N_1$ and $N_2$, so that they decay completely into $N_3$ and $\sigma$. Shortly after, the temperature falls below the mass of $N_3$ and its density begins to redshift to $a^{-3}$, while the light $\sigma$ scalar is diluted as $a^{-4}$. Later, $N_3$ decays and its energy is transferred into Standard Model particles at a temperature $T_D$. From this we find:
\begin{equation}
\frac{a_{N_3}}{a_D}
=
\left[
\frac{4}{7}
g_*^{SM}(T_D)
\right]^{1/3}
\left(\frac{T_D}{M_{N_3}}\right)^{4/3}.
\end{equation}
where $a_{N_3}$ denotes the scale factor when $T=M_{N_3}$, and $g_*^{SM}$ is the number of relativistic degrees of freedom of the Standard Model. The total entropy density after $N_3$ decays includes the contribution of the light $\sigma$ scalar and of the Standard Model degrees of freedom, giving:
\begin{eqnarray}
s(T_D)&=& s_\sigma(M_{N_3})\left({a_{N_3}\over a_D}\right)^3 +s_{SM}(T_D)\nonumber\\
&=&s_{SM}(T_D)\left[1+{4\over 7}{T_D\over M_{N_3}}\right]
\end{eqnarray}
The ratio between the comoving entropy densities at $T=M_{N_s}$ and $T=T_D$ is thus given by:
\begin{eqnarray}
\Delta_s^{-1}= {s(T_D)a_D^3\over s(M_{N_3})a_{N_3}^3}=
\frac7{11}
\left(
1+\frac47\frac{T_D}{M_{N_3}}
\right)
\frac{M_{N_3}}{T_D}~,
\end{eqnarray}
where we note that, for $T\gtrsim M_{N_3}$,  $s=s_\sigma+s_{N_3}= {11\over4}s_\sigma$. The factor $\Delta_s^{-1}$ thus corresponds to the increase in the comoving entropy when $N_3$ decays. Hence, the yield of the axion, $n_a/s$, and consequently its abundance, are reduced by $\Delta_s$ after $N_3$ decays. In the example given in the figures above, lukewarm inflation ends at a temperature of $\simeq 450$ MeV, which gives an upper bound on the mass of $N_3$. Saturating this bound and taking $T_D=5$ MeV yields $\Delta_s^{-1}\simeq 58.1$. 

\begin{figure}[!h]
    \centering
    \includegraphics[width=1\linewidth]{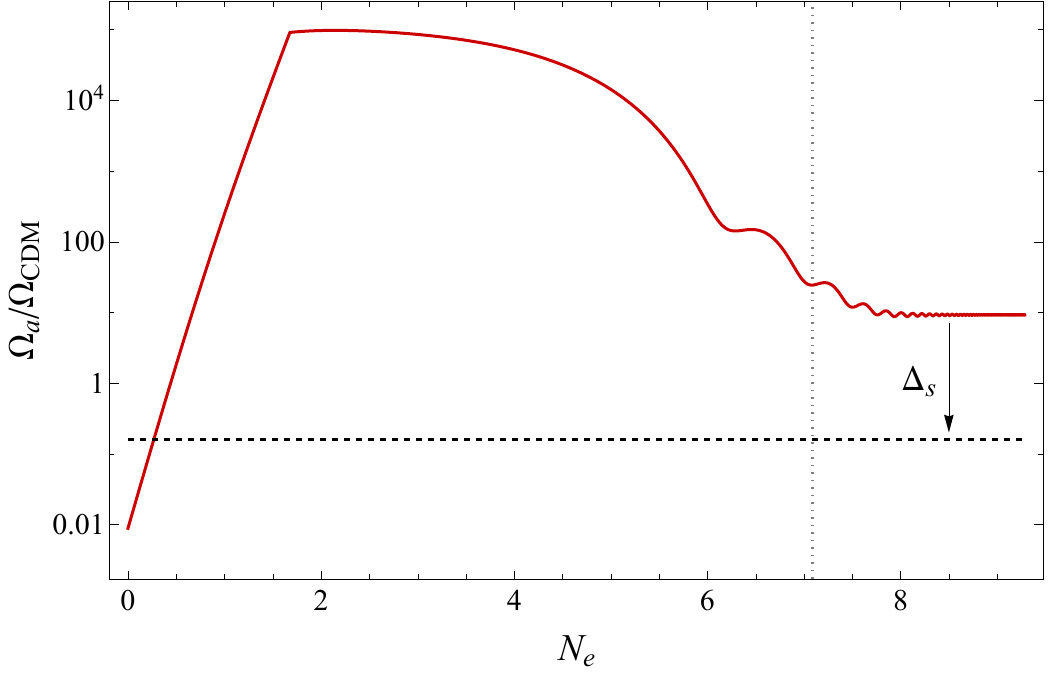}
    \caption{Evolution of the QCD axion abundance during lukewarm inflation, for the same parameter choices as in the previous figure. The dotted vertical line marks the end of the lukewarm inflation stage ($T_E=450$ MeV). Note that the axion abundance is further reduced by the entropy injection from the decay of the lightest right-handed neutrinos after lukewarm inflation. The dashed horizontal line indicates the minimum present axion abundance considering $T_D=5$ MeV.}
    \label{fig: axionomega1}
\end{figure}

In Figure \ref{fig: axionomega1}, we show the evolution of QCD axion abundance, $\Omega_a=\rho_a/\rho_c$, normalized to the abundance of cold dark matter, $\Omega_{CDM}h^2\simeq 0.12a^{-3}$ (setting $a(t_0)=1$). We see that in this example the QCD axion is still overabundant at the end of the lukewarm inflation stage by a factor $\sim 27$, implying $M_{N_3}\gtrsim 200$ MeV to bring its abundance below the measured cold dark matter value. In this example the QCD axion accounts for a fraction of dark matter, despite its large decay constant and initial misalignment angle.

The lukewarm inflation stage dilutes not only the QCD axion density but also any other thermal relics generated earlier in the cosmic history. This includes, in particular, the baryon asymmetry. Given that most baryogenesis scenarios proposed in the literature operate at temperatures above the GeV scale considered in this work, we envisage a scenario where a large baryon asymmetry was produced before lukewarm inflation, e.g.~through the Affleck-Dine mechanism \cite{Affleck:1984fy, Dine:1995kz}, and then subsequently diluted by the latter period, alongside the additional injection of entropy from $N_3$ decays.   

\begin{figure}[!h]
    \centering
    \includegraphics[width=1\linewidth]{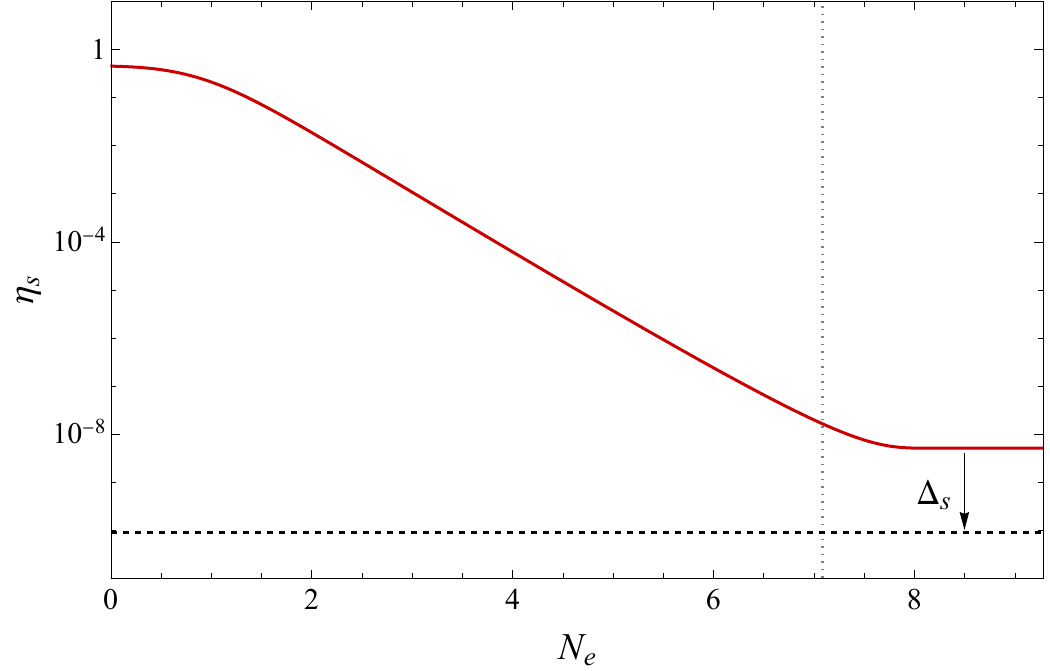}
    \caption{Evolution of the baryon-to-entropy ratio during lukewarm inflation, for the same parameter choices as in the previous figures and an initial value $\eta_{s}^i \simeq 0.46$ (generated before lukewarm inflation). The dotted vertical line marks the end of the lukewarm inflation stage ($T_E=450$ MeV). Note that the baryon-to-entropy ratio is further reduced by the entropy injection from the decay of the lightest right-handed neutrinos after lukewarm inflation. The dashed horizontal line indicates the minimum present value of $\eta_s$ considering $T_D=5$ MeV, which coincides in this case with the measured value.}
    \label{fig: eta1}
\end{figure}

In Figure \ref{fig: eta1}, we show the evolution of the baryon-to-entropy ratio, $\eta_s=n_B/s$, for the same parameters as in Figure \ref{fig: totaldyn1}, for an initial value $\eta_{s}^i \simeq 0.46$. Since $n_B\propto a^{-3}$ (assuming, of course, no baryon number violation at this stage) and the entropy density decreases slowly during the lukewarm inflation stage, the baryon-to-entropy ratio is greatly reduced during this period, considerably more than the QCD axion that is only efficiently diluted close to the end of the secondary inflation period. Note that the additional entropy injection from $N_3$ decays after lukewarm inflation will also dilute the baryon-to-entropy ratio, yielding in this illustrative example a final value coinciding with the presently measured one, $\eta_{s,0}\simeq 8.66\times10^{-11}$ \cite{Planck:2018vyg}.

\begin{figure}[!ht]
    \centering
    \includegraphics[width=1\linewidth]{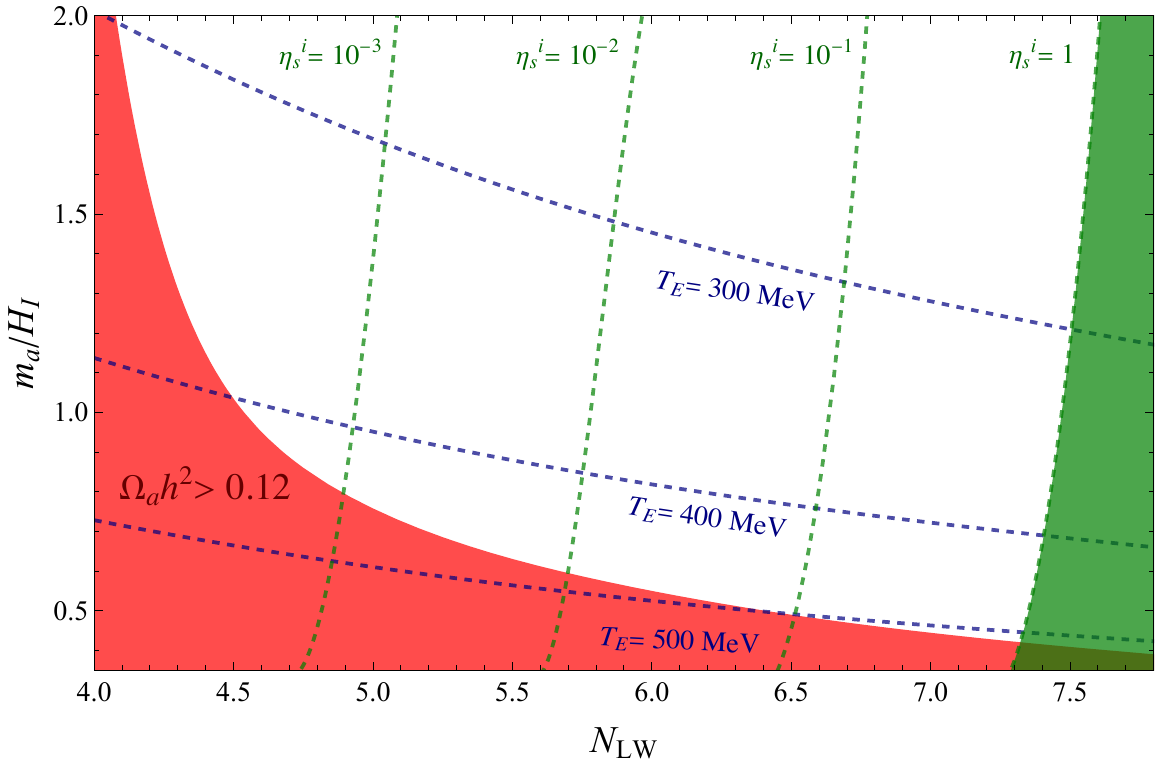}
    \includegraphics[width=1\linewidth]{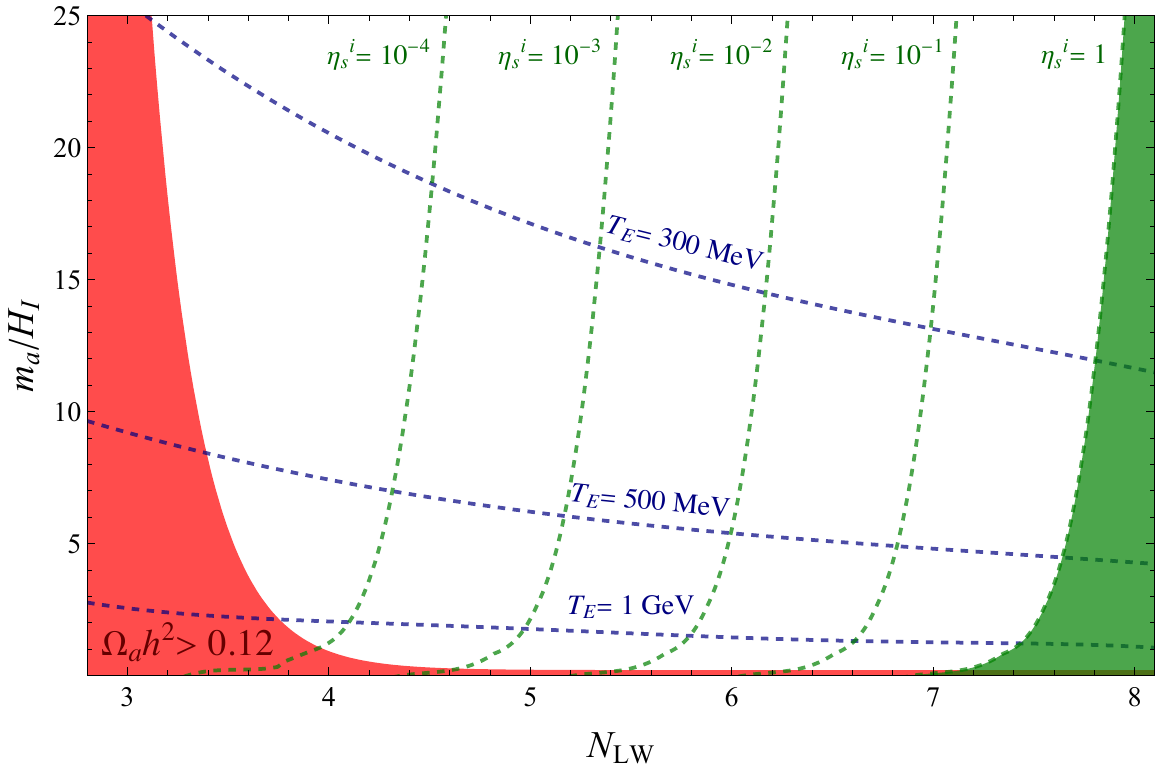}
    \caption{Parameter space $(N_{LW},m_a/H_I)$ with $\theta_i= \pi/\sqrt{3}$, $C_T\simeq 10^{-3}$ for decay constant $f_a = 10^{16}$ GeV (top) and $f_a = 10^{15}$ GeV (bottom). In the red shaded region the axion is overabundant and in the green shaded region the baryon asymmetry is below the current observed value for an initial value $\eta_s^i \sim 1$. The green dashed lines represent the current observed baryon asymmetry for different initial values. The purple dashed lines represent several different temperatures at the end of inflation that we assume to match the mass of the right-handed neutrino $M_{N_3}$.}
    \label{fig: moneyplot1}
\end{figure}

In Figure \ref{fig: moneyplot1} we show the results obtained for the final abundance of the QCD axion, for $\theta_i=\pi/\sqrt{3}$, and the final baryon-to-entropy ratio by numerically solving the relevant dynamical equations system during lukewarm inflation for different values of the inflationary scale, i.e., the ratio $m_a/H_I$, and the duration $N_{LW}$. The final abundance considers the maximum entropy injection factor from $N_3$ decays, with $M_{N_3}=T_E$. We fix $C_T=10^{-3}$, although our results do not depend on this choice, since the dynamics of the axion depends only on the scale and duration of the lukewarm inflation period. We consider two different values of the axion decay constant, $f_a=10^{16}$ GeV (top) and $f_a=10^{15}$ GeV (bottom). 

In the shaded red regions of the parameter space, the QCD axion abundance exceeds the measured abundance of cold dark matter, and we see that viable models have longer periods of lukewarm inflation for larger values of the Hubble scale (smaller $m_a/H_I$). This is expected since, for larger inflationary scales, the axion begins to oscillate later, thus requiring a larger number of e-folds. In the non-shaded white region, the QCD axion accounts for only a fraction of dark matter. 

In this figure, we also plot contours of the initial baryon-to-entropy ratio leading to the presently measured value (dashed green curves). The shaded green region is thus excluded since it would yield a too small baryon asymmetry. This means that viable models of lukewarm inflation for GUT scale axion decay constants require roughly 4-8 e-folds of accelerated expansion.

The plots in Figure \ref{fig: moneyplot1} also show contours for the temperature at the end of the lukewarm inflation period, which we recall sets an upper bound on the right-handed neutrino masses. We see that larger values of $f_a$, i.e.~smaller axion masses, require lower temperatures and hence lighter right-handed neutrinos. For $f_a\lesssim 10^{16}$ GeV, we find viable QCD axion scenarios with right-handed neutrino masses $\gtrsim 100$ MeV, such that they decay into Standard Model particles before BBN. For larger values of $f_a$, however, lukewarm inflation would have to occur at temperatures below 100 MeV, and viable QCD axion scenarios would require tuning the initial misalignment angle.


\section{Observational consequences}\label{sec:Observ}

We have shown in the previous section that a late stage of lukewarm inflation may efficiently dilute the abundance of the QCD axion, in scenarios with an axion decay constant close to the GUT scale, at the same time also bringing a large baryon-to-entropy ratio down to the measured value. Let us now explore some of the observational consequences of lukewarm inflation, particularly those that are specific to the particle physics setup and inflationary scale discussed in this work. The most generic consequence, the enhancement of the small-scale curvature power spectrum and the consequent genesis of primordial black holes and scalar-induced gravitational waves, is addressed in a companion paper, and here we will simply lay the ground for this discussion.

\subsection{Dark radiation} \label{Sec:Neff}

A crucial aspect of the proposed particle physics setup is the additional light scalar $\sigma$, one of the decay products of the two heaviest right-handed neutrinos, ${N_{1,2}\rightarrow N_3+\sigma}$. Although all right-handed neutrinos necessarily decay into light Standard Model degrees of freedom, the corresponding decay width is too small to maintain the thermal bath close to equilibrium during lukewarm inflation for right-handed neutrino masses of a few hundred MeV. The additional decay channel is therefore crucial for a consistent realization of lukewarm inflation at such low temperatures.

Given the sub-eV mass of the ``lukewarm inflaton'' field $\phi$, it is natural to assume that the only other scalar coupled to the right-handed neutrinos is also comparably light. This means that after lukewarm inflation ends, and the temperature falls below the right-handed neutrino masses, the $\sigma$ scalar behaves as dark radiation, contributing to the number of relativistic degrees of freedom present at BBN and recombination. 

In the post-lukewarm inflation universe, the energy balance is dominated by $N_3$ and $\sigma$, since $N_1$ and $N_2$ decay as soon as the temperature falls below their mass. As discussed above, $N_3$ becomes non-relativistic after lukewarm inflation, and naturally becomes the dominant species. This means that Standard Model degrees of freedom eventually dominate the radiation bath once $N_3$ decays, before BBN takes place, with $\sigma$ giving only a small contribution to the number of relativistic species.

The amount of dark radiation is conventionally expressed in the literature as an effective number of light neutrino species, such that:
\begin{align}
    \frac{\rho_\sigma}{\rho_\gamma} = \frac{7}{8}\left(\frac{4}{11}\right)^{4/3}\Delta N_{eff}~.
\end{align}

When the lightest right-handed neutrino decays into the Standard Model, we have:
\begin{equation}
{\rho_\sigma\over \rho_{SM}}(T_D)=\left({4\over 7}{ T_D\over M_{N_3}}\right)^{4/3}g_*^{SM}(T_D)^{1/3} 
\end{equation}
Since for $T<T_D$ entropy is conserved in the Standard Model sector, $g_{*S}^{SM}(T) T^3 a^3 = g_{*S}^{SM}(T_D) T_D^3 a_D^3$, and using $\rho_\gamma=2\rho_{SM} / g_*^{SM}$, we find after some algebra:
\begin{equation}
\Delta N_{eff} ={4\over 7}\left[{11\over 7} {T_D\over M_{N_3}}g_{*s}^{SM}(T)\right]^{4/3}~.
\end{equation}
Using $g_{*s}\simeq 3.91$ for the entropy degrees of freedom after electron-positron annihilation, we then find ${\Delta N_{eff}< 0.047}$ for $T_D=5$ MeV and right-handed neutrinos heavier than 200 MeV. This is within the current observational bounds \cite{AtacamaCosmologyTelescope:2025nti} and could potentially be probed with the future Simons Observatory \cite{SimonsObservatory:2018koc,SimonsObservatory:2025wwn}.

\subsection{Enhancement of small-scale curvature perturbations}

As mentioned earlier, interactions between the inflaton scalar field and the ambient nearly-thermal radiation bath result not only in thermal friction but also in thermal fluctuations. These are sourced by the approximately Gaussian white-noise term in Eq.~(\ref{langevin}), with a variance related to the thermal friction coefficient $\Upsilon$ via the fluctuation-dissipation relation. This means that in lukewarm inflation, as for warm inflation scenarios in general, inflaton perturbations are sourced by both quantum and thermal effects, the latter being dominant for $T\gtrsim H$ \cite{DeOliveira:2001he,Graham:2009bf,Bastero-Gil:2019rsp}. Thermal friction also damps the evolution of field fluctuations and associated curvature perturbations, which ``freeze out'' even before becoming super-horizon in the strong dissipation regime, $Q\gg 1$. Moreover, the temperature dependence of the dissipation coefficient $\Upsilon(T)$ introduces a non-linear interplay between the inflaton and radiation perturbations that, for $d\Upsilon/dT>0$, results in a growing mode \cite{Graham:2009bf, Bastero-Gil:2011rva}. The dimensionless curvature power spectrum takes the analytical expression \cite{Ramos:2013nsa}:
\begin{align}
    \Delta_\mathcal{R}^2 =\frac{V_*(1+Q_*)^2}{24\pi^2M_p^4\epsilon_\phi}\left(1+2n_*+\frac{T_*}{H_*}\frac{2\sqrt{3}\pi Q_*}{\sqrt{3+4\pi Q_*}}\right)G(Q_*),
\end{align}
where all quantities labeled with $*$ are evaluated at horizon crossing and $G(Q)$ is a function computed numerically that takes into account the coupling between the inflaton and radiation perturbations \cite{Bastero-Gil:2011rva,Montefalcone:2023pvh,Rodrigues:2025neh}. In the strong dissipation regime, $G(Q)\simeq \alpha Q^\beta$, and using the slow-roll equations for $\Upsilon=C_T T$ the dimensionless power spectrum can be rewritten in the form:
\begin{align}
    \Delta_\mathcal{R}^2\simeq \left(\frac{\rho_R}{12\pi^2 M_p^4\epsilon_H^2}\right)\left(\frac{6\sqrt{3}\pi\alpha}{\sqrt{4\pi}C_T}\right)Q_*^{5/2+\beta},
\end{align}
The slow-roll equations for the inflaton and radiation equations also yield:
\begin{align}
    \left(\frac{T}{H}\right)^2 = \frac{3\epsilon_H}{2C_R}\left(\frac{M_p}{T}\right)^2~,
\end{align}
where we used $\rho_R = C_R T^4$ and $3H^2M_P^2 \simeq  V$. Then
\begin{align}
    \Delta_\mathcal{R}^2\simeq \frac{\sqrt{3\pi}}{12\pi^2}\frac{\alpha\epsilon_H^{\beta/2-3/4}}{3^{1/4+\beta/2}2^{5/4+\beta/2}}{C_T^{3/2+\beta_j}\over C_R^{1/4+\beta/2}}\left(\frac{T}{M_p}\right)^{3/2-\beta}.
\end{align}

Using the numerical package WI2easy \cite{Rodrigues:2025neh}, we have obtained $\alpha\simeq 5\times10^{-4}$ and $\beta\simeq 3$, resulting in
\begin{align}
    \Delta_\mathcal{R}^2\simeq 0.004\epsilon_H^{3/4}\left(\frac{10}{g_*}\right)^{7/4}\left(\frac{C_T}{10^{-5}}\right)^{9/2}\left(\frac{1\text{ GeV}}{T}\right)^{3/2}~.\label{Eq:powerspectrum}
\end{align}
This is applicable to the range of comoving scales that become super-horizon during the lukewarm inflation period. We thus see that a generic prediction of a secondary period of warm inflation is a large enhancement of the curvature perturbations power spectrum on small scales. In fact, Eq.~(\ref{Eq:powerspectrum}) yields a temperature-dependent upper bound on the value of the dissipation constant $C_T$ (and hence on the ratio of the Yukawa couplings $g$ and $h$ in our model) such that $\Delta_\mathcal{R}^2<1$. This corresponds to $C_T\lesssim 10^{-5}$ in the present scenario for QCD axion dilution, with a lukewarm inflation at temperatures below the GeV scale.

Notably, the amplitude of the curvature perturbations generated during (luke)warm inflation decreases with the temperature. Hence, large curvature perturbations are an intrinsic feature of a secondary period of warm inflation occurring at late times. 

While the amplitude of the power spectrum is independent of the form of the scalar potential, the latter affects the associated spectral index, which is given by:
\begin{align}
    n_s-1 = \frac{d\log \Delta_{\mathcal{R}}^2}{d\log k} \simeq \frac{d\log \Delta_{\mathcal{R}}^2}{dN_e} &= \frac{3}{4}\frac{d\log\epsilon_H}{dN_e}-\frac{3}{2}\frac{d\log T}{dN_e}.
\end{align}
Using the relations
\begin{align}
&\epsilon_H\simeq\frac{\epsilon_\phi}{1+Q}~~,\\
&\frac{d\log\epsilon_H}{dN_e} =\frac{d\log\epsilon_\phi}{dN_e}-\frac{d\log(Q+1)}{dN_e}~,\\
&\frac{d\log\epsilon_\phi}{dN_e} = -\frac{2}{1+Q}\left(\eta_\phi-2\epsilon_\phi\right)~,\\
&\frac{d\log T}{dN_e} = \frac{d\log Q}{dN_e}-\epsilon_H~,
\end{align}
and the slow-roll equations \eqref{slow-roll}, the spectral index can be written, in the strong dissipation regime $Q\gg1$, as
\begin{align}
    n_s-1\simeq \frac{3}{10Q}\left(6\epsilon_\phi-2\eta_\phi\right)~.
\end{align}
Curiously, note that this involves the same combination of slow-roll parameters as in single-field (cold) inflation models, albeit with the opposite sign. This means that scalar potentials yielding red-tilted spectra in cold inflation scenarios result in blue-tilted spectra in (luke)warm inflation. This is, on the one-hand, the case of the quadratic potential that we considered in our illustrative example of the lukewarm inflation dynamics. A hybrid-like potential $V(\phi)= V_0 + m^2\phi^2/2$ would, on the other hand, lead to a red-tilted spectrum on small-scales. The freedom in the choice of the scalar potential associated with the gauge-invariant nature of the inflaton field in the Warm Little Inflaton setup thus means that either case is possible.

The substantial enhancement of small-scale curvature perturbations has two natural observational consequences $-$ the formation of primordial black holes and the generation of a stochastic background of secondary gravitational waves. These are, as shown above, a generic feature of a secondary lukewarm inflation stage, not necessarily linked to the particular implementation considered in this work or to the inflationary scales and temperatures required for an efficient dilution of the QCD axion. As such, we discuss these observational consequences in a separate companion paper \cite{us}.

An aspect that is particular to the lukewarm inflation scenario discussed in the present paper is the occurrence of a subsequent early-matter era, dominated by the lightest right-handed neutrino before its decay into Standard Model particles. This may have an impact on both primordial black holes and gravitational waves, particularly in the former since the threshold for black hole formation is much lower in a universe dominated by pressureless matter. Primordial black hole formation in matter-domination is, however, presently plagued with some uncertainty given the number of effects that become relevant in the absence of ambient pressure (e.g.~angular momentum, anisotropic collapse). We will therefore leave a detailed analysis of the effects of the early-matter era for a future dedicated work.

\section{Conclusions}\label{sec:concl}

In this work we have shown that a secondary period of warm inflation, which we denote as lukewarm inflation, lasting less than 10 e-folds, may efficiently dilute the abundance of the QCD axion, leading to viable scenarios with axion decay constants close to the GUT scale, $f_a=10^{15}-10^{16}$ GeV,

We have proposed a novel particle physics setup to implement the ``Warm Little Inflaton'' scenario at low temperatures  based on the collective spontaneous breaking of a gauge $U(1)$ symmetry that may be identified with lepton number. The ``lukewarm inflaton'' therefore interacts with a nearly thermal bath of three right-handed neutrinos and an additional light scalar, resulting in a large thermal friction that sustains both the slow-roll field trajectory and the temperature of the thermal bath. Two of the right-handed neutrinos interact directly with the inflaton and decay primarily into the third (and lightest) right-handed neutrino and the light scalar.

While this setup could potentially be used to realize lukewarm inflation (or, in fact, a primary period of warm inflation) at arbitrary temperatures, here we have focused on scenarios where lukewarm inflation occurs around a few hundred MeV, for which the Hubble scale is around the QCD axion mass of $0.1-1$ neV. 

At the start of lukewarm inflation, the Standard Model degrees of freedom are quickly diluted away and the axion mass reaches its zero-temperature value. The axion field is therefore critically damped and evolves slowly, until eventually the Hubble scale drops sufficiently for it to begin oscillating and thus behaving as cold dark matter. We have shown that this initial critical damping may substantially reduce the field value at the onset of oscillations. Alongside the inflationary dilution in the oscillating phase and the nearly constant entropy density during lukewarm inflation, this can dramatically decrease the QCD axion yield, $n_a/s$, and therefore its present abundance.

In addition, in the proposed particle physics setup the lightest right-handed neutrino tends to dominate the energy balance after the lukewarm inflation stage, leading to an early matter-dominated epoch. It eventually decays into Standard Model degrees of freedom, through the standard Yukawa couplings that also explain light neutrino masses through the seesaw mechanism. This leads to further entropy injection and dilution of the QCD axion abundance (although most of the dilution occcurs during lukewarm inflation). Note that requiring that this decay occurred before BBN limits the right-handed neutrino masses to be greater than $\sim 100$ MeV, depending on the form of the neutrino mass matrix, which constrains the available parameter space. However, we have identified viable parametric regions for axion decay constants up to $10^{16}$ GeV, with no fine-tuning of the initial misalignment angle, beyond the reach of other scenarios proposed in the literature. 

Note that scenarios where the Peccei-Quinn symmetry is broken close to the GUT scale are not only theoretically well-motivated but also appealing from the phenomenological perspective, given the current CMB bounds on axion isocurvature perturbations, which limit the scale of the primary inflation period (during which CMB scales become super-horizon) to ${H^*\lesssim2.8 \times10^{11}\, \mathrm{GeV}\, \theta_i \left(f_a/10^{16}\mathrm{GeV}\right)}$ if the QCD axion accounts for all the dark matter \cite{OHare:2024nmr}. In fact, there are no axion isocurvature modes if the primary inflation period is also a warm inflation stage and the Peccei-Quinn transition occurs only after CMB scales cross the horizon \cite{Rosa:2021gbe}.

The lukewarm inflation stage dilutes, of course, not only the QCD axion but also any other thermal or non-thermal relic generated after the first inflation period. This includes, in particular, the baryon asymmetry, and even though there are scenarios that could in principle generate a baryon/lepton asymmetry after lukewarm inflation, we have shown that the measured value could be obtained if a relatively large baryon-to-entropy ratio is produced before the secondary inflation stage. An interesting feature of our proposed setup is the fact that baryon number and other cold relics are more efficiently diluted than the QCD axion number density, given the latter's initial critically damped evolution.

Our proposed scenario also leads to dark radiation, in the form of the light scalar resulting from the decay of the heaviest right-handed neutrino. This contributes to the number of relativistic species at both BBN and recombination, yielding $\Delta N_{eff}\lesssim 0.05$ within the reach of future CMB experiments. While the ``lukewarm inflaton'' in our setup is also a relatively light scalar, with a sub-eV mass, it contributes to cold dark matter as an oscillating scalar field, being asymptotically stable as shown in \cite{Rosa:2018iff} for the generic Warm Little Inflaton construction. Its contribution is, however, sub-dominant compared to the QCD axion, given the efficiency of thermal friction in transferring its energy density into the radiation bath during and at the end of the lukewarm inflation stage, as illustrated in Figure \ref{fig: totaldyn1}.

Lukewarm inflation enhances, in general, the amplitude of primordial curvature perturbations on small-scales, despite the low inflationary Hubble scale. This is due to both the thermal nature of scalar inflaton fluctuations and their coupling to fluctuations in the temperature of the thermal bath, which result in a growing mode. A natural outcome is thus the formation of (sub-solar) primordial black holes and associated stochastic gravitational wave background, which we analyze in detail in a companion paper \cite{us}. While this is a generic feature of a secondary warm inflation stage, the proposed realization with temperatures around a few hundred MeV may leave particular imprints on the primordial black hole mass function, namely due to the early matter-dominated epoch that necessarily follows lukewarm inflation in this case, and which we plan to investigate in more detail in the future.

\acknowledgments
This work was supported by national funds by FCT - Fundação para a Ciência e Tecnologia, I.P., through the research projects with DOI identifiers 10.54499/UID/04564/2025 and by the project 10.54499/2024.00252.CERN funded by measure RE-C06-i06.m02 – “Reinforcement of funding for International Partnerships in Science, Technology and Innovation” of the Recovery and Resilience Plan - RRP, within the framework of the financing contract signed between the Recover Portugal Mission Structure (EMRP) and the Foundation for Science and Technology I.P. (FCT), as an intermediate beneficiary. P. B. F. was supported by the FCT - Fundação para a Ciência e Tecnologia, I.P. fellowship SFRH/BD/151475/2021 with DOI identifier 10.54499/SFRH/BD/151475/2021. ATM was supported by the Slovenian Quantum Science Hub co-funded by the Marie Skłodowska-Curie Actions programme (GA-101177446) and the Slovenian Research and Innovation Agency (ARIS), contract number 5110-18/2025-5.

\bibliography{main_draft}

\end{document}